\documentclass[11pt, a4paper]{article}
\usepackage[utf8]{inputenc}
\usepackage{amsmath}
\usepackage[sorting=none,style=numeric-comp,url=false]{biblatex}
\usepackage{commath}
\usepackage{geometry}
\usepackage{graphicx}
\usepackage{mathtools}
\usepackage{setspace}
\usepackage{bm}
\usepackage{etoolbox}
\usepackage{braket}
\usepackage{indentfirst}
\usepackage{caption}
\usepackage{float}
\usepackage{authblk}
\AtEveryBibitem{
 \clearfield{urlyear}
 \clearfield{urlmonth}}
\title{Kerr-NUT Thermodynamics:\\ Misner Strings and Nut Charges}

\author[a]{Mohamed Tharwat \thanks{oweg@aucegypt.edu}}
\author[a,b]{Adel Awad \thanks{a.awad@sci.asu.edu.eg}}

\affil[a]{\footnotesize \it Centre for Theoretical Physics, the British University in Egypt, El Sherouk City 11837, Egypt}
\affil[b]{\footnotesize \it Department of Physics, Faculty of Science, Ain Shams University, Cairo 11566, Egypt}
\date{July 2026}

\begin{document}
\maketitle
\begin{abstract}
 Using the thermodynamic approach introduced in e-Print: 2206.09124 [hep-th] to the case of Kerr-NUT solution, we demonstrate the existence of a new conserved quantity that is dual to the angular momentum, in the same way that the nut charge "n" is dual to the mass "m". This quantity exists regardless of the presence of the Kerr rotation parameter $a$, but only affects its thermodynamics when $a\neq 0$. The thermodynamic quantities calculated together with the two charges $N_{c}=cn$ and $N_{n}=n$ lead to consistent thermodynamic relations. That is, the first law, Smarr's and Gibbs-Duhem relations are all satisfied. 
\end{abstract}
\newpage
\section{Introduction}

The Taub-NUT solution was first discovered in 1951 by Taub \cite{taub_empty_1951}. At that time, it was presented in coordinates that only covered the time-dependent part of the spacetime. Later, Newman, Unti, and Tumbrino presented the metric in a different form, which generalizes the Schwarzchild spacetime \cite{newman_emptyspace_1963}. 
Another class of solution that generalizes the Taub NUT solution is the Kerr-NUT solution \cite{demianski_combined_1966}, where it includes a rotation parameter $a$ in addition to the nut and mass parameters. It is intriguing to observe that this solution carries an angular momentum due to an additional parameter which we call "nut-associated" parameter, $c$, in addition to the rotation parameter $a$. The parameter $c$ exists in the nonrotating solution as well, but most authors set it to vanish. It is known that different values of $c$ correspond to changing the position of Misner string! In this work, we will keep $c$ nonvanishing, in order to show its role in constructing a consistent thermodynamics for the Kerr-NUT solution.

The study of the solution and its thermodynamics was largely shaped by Misner's work, which demonstrated the existence of string-like singularity. This string can be removed upon enforcing a periodicity condition on the timelike coordinate \cite{misner_flatter_1963}. However, this has various consequences for the thermodynamics of the spacetime and the geodesic completeness of the spacetime. 

The periodicity condition significantly constrains the thermodynamics of the solution, through relating the the nut charge to horizon radius, $T(r_h)=\frac{1}{\beta}=\frac{1}{8 \pi n}$. Since the degrees of freedom are reduced, we are unable to form a first law that is of full cohomogeneity. Another major consequence is the possibility of having a negative entropy where $S\neq\frac{A}{4}$ \cite{johnson_extended_2014}. Much of the earlier work was concerned with the Euclidean solution due to adherence to the periodicity condition\cite{hawking_gravitational_1999,hawking_nut_1999,hunter_action_1998,chamblin_large_1999,awad_higher_2006,fatibene_entropy_2000,johnson_thermodynamic_2014}.

Although, Misner's solution successfully treats the singular behaviour of the timelike coordinate \cite{misner_flatter_1963}, it introduces closed timelike curves all over the spacetime. The causal pathology without the condition was shown to be less perverse than initially considered \cite{clement_rehabilitating_2015}. 
Additionally, the geodesic completeness of the maximally extended spacetime is also affected. If the periodicity is not imposed, the spacetime can be maximally extended into a geodesically complete spacetime. It was shown by Cl\'ement et al. that the Misner strings are completely transparent to the geodesics passing through them \cite{clement_rehabilitating_2015}. 

These developments alleviated some of the concerns over the interpretation of the metric as a meaningful physical solution to the Einstein field equations. It also opened the possibility of investigating the Taub-NUT thermodynamics without imposing the periodicity condition\cite{abbasvandi_thermodynamics_2021,bordo_misner_2019,hennigar_thermodynamics_2019,bordo_first_2019,bordo_thermodynamics_2020,awad_lorentzian_2022}. The earliest work that discussed this possibility and gave an alternative interpretation of the spacetime was that of Bonnor, where he treated the strings as a singular source of angular momentum \cite{bonnor_new_1969}.

Additionally, the Taub-NUT-AdS solutions have attracted interest due to their interesting phase behaviour \cite{xiao_thermodynamical_2021,awad_dyonic_2023,ballon_thermodynamics_2019,albarqawy_dyonic_2025-1,tharwat_dyonic_2024}. The phase structure has been studied for several horizon geometries and ensembles, with most of the work focusing on the extended phase space. The dyonic spherical solution is known to carry an additional critical point when compared to the $n=0$ case \cite{awad_dyonic_2023,tharwat_dyonic_2024,ballon_thermodynamics_2019}. The dyonic solutions with hyperbolic and flat horizon geometries were shown to admit a first order phase transition for the Einstein-Maxwell action \cite{tharwat_dyonic_2024}. Working outside the extended phase space, the flat horizon geometry admits more than one critical point \cite{albarqawy_dyonic_2025-1}. 

The Taub-NUT solution is considered as a gravitational dyon since the nut parameter can be interpreted as a "dual mass" in analogy to electromagnetic duality \cite{ortin_gravity_2015,griffiths_taubnut_2010}. This is motivated by the duality between the electric and magnetic charges in $d=4$ where the Hodge dual of the two form used to calculate the electric charge can be used to calculate the magnetic charge. Another reason is the analogy between the Misner strings in the gravitational solution and the Dirac string in electromagnetism. 

Similar to the physical interpretation of the metric, the thermodynamics of the solution are non-trivial; different thermodynamic treatments have been suggested \cite{bordo_misner_2019,awad_lorentzian_2022,mann_complement_2021,bordo_thermodynamics_2020,abbasvandi_thermodynamics_2021,corral_nonconformally_2026,ciambelli_topological_2021}. For the Kerr-NUT solution in particular, more than one approach exists in the literature \cite{bordo_first_2019,rodriguez_first_2021,bordo_thermodynamics_2020,frodden_first_2022,yang_first_2023,liu_thermodynamics_2022,corral_electric-magnetic_2024}. These approaches differ significantly from our work for several reasons, we discuss these differences below.

Here it is important to briefly highlight the main differences between different thermodynamics approaches in the literature. The most significant demarcator between formulations is the addition of a work term to the first law that depends on $n$. One main difference between these approaches is whether the formulation attributes an entropy to the Misner strings \cite{bordo_misner_2019,abbasvandi_thermodynamics_2021}. For cases where the entropy receives no contribution from strings, different identifications of nut charges and potentials exist \cite{awad_lorentzian_2022,bordo_misner_2019,bordo_thermodynamics_2020}.

Upon applying the above periodicity condition, the first law is inevitably of reduced cohomogeneity since no additional work term can be considered. To construct consistent thermodynamics where no $n$-dependant term is present, we must reduce the parameter space to one where the nut parameter is a multiple of another thermodynamic quantity. For example, in one analysis of the Kerr-NUT solution the identification was chosen as $n=\alpha m$ where $\alpha$ is not fixed a priori \cite{frodden_first_2022}. It was not allowed to vary ($\delta\alpha =0$ is assumed). The authors in this work used the surface charge method to determine the charges.

For approaches where a charge $N$ is associated with $n$, the method used to identify the nut charge varies. There are several different identifications in the literature. Our treatment follows from the interpretation of the nut parameter $n$ as the "magnetic-type" mas dual to the "electric-type" mass \cite{dowker_nut_1974,awad_lorentzian_2022}. One approach appeals to the presence of a Killing horizon around the strings. They associate the surface gravity of each string with a nut potential $\psi^{'}_{\pm}$ where a conjugate nut charge $N^{'}_{\pm}$ is defined for each potential \cite{bordo_misner_2019,bordo_first_2019}. Another approach is centred around the introduction of secondary hair $J_{n}=mn$ which corresponds to the mass of a magnetic monopole in $5$ dimensions. 

We prefer not to use the surface gravity of the strings as thermodynamic potentials for more than one reason. The main qualm is the lack of justification for the conservation of the associated nut charges $N^{'}_{\pm}$ \cite{bordo_first_2019}. Additionally, interpreting $\psi^{'}_{\pm}$ as a surface gravity that correspond to temperatures, imply the existence of a multi-temperature system which leads to restrictions on the thermodynamics that are similar to the Misner periodicity condition \cite{bordo_first_2019}. 

We also disagree with the introduction and use of $J_{N}=mn$ \cite{wu_thermodynamical_2019,wu_consistent_2022}. The inclusion of the this term leads to a first law that depends on $(3)$ thermodynamic parameters $A_{h},J_{N}$, and $N$. However, this exceeds the dimensions afforded by the $(2)$ independent parameters $r_{h}$ and $n$. Consequently, it is impossible to vary the three thermodynamic parameters present in \cite{wu_thermodynamical_2019,wu_consistent_2022} independently which results in a first law that is not unique.

Here, we are going to study the Kerr-NUT solution using an approach that relies on the use of conserved thermodynamic quantities to construct the first law. The first conserved charge is $N_{n}=n$ and is equivalent to the non-rotating case. For the second charge we introduce $N_{c}=cn$, which we show to be a conserved quantity. This is demonstrated through constructing a dual quantity to the angular momentum, which implies the conservation. I.e., similar to the mass, the angular momentum can be calculated using $\star d\chi$ where $\chi$ is the rotational Killing vector $\partial_{\phi}$. We show that the integral over its Hodge dual results in the conserved quantity
\begin{equation}
\frac{-1}{8\pi}\int_{S^{2}_{r\rightarrow\infty}} d\chi=2cn^{2},
\end{equation}
and that its conservation implies that $cn$ is a conserved quantity.

The present work is broken down as follows. In section 2, we use Komar integrals to calculate the thermodynamic charges and show that they satisfy the first law and the Smarr relation. The identification of the nut charges $N_{\pm}$ is chosen in a way that makes it easy to compare our formulation with other works in the literature. Section 3 expands on the possible choices for the nut charges as well as their implications. We introduce the thermodynamic charges $N_{n}$ and $N_{c}$ and demonstrate that they are conserved quantities. In Section 4, we calculate the Euclidean action and use it to show that our charges also satisfy the Gibbs-Duhem and differential Gibbs energy equation. This is used as self-standing verification of our approach. 

\section{The First Law}
In this section, we obtain the mass and thermodynamic charges using Komar integrals. This method has the advantage of manifestly satisfying the Smarr relation where the relation is simply a consequence of the integrals closing over the designated spacelike boundaries. The metric for the Kerr-NUT solution is given by 

\begin{equation}
\begin{aligned}
    ds^{2}=&-\frac{\Delta_{r}}{\Sigma}\left(dt+\left(2n(\cos{\theta} +c )-a\sin^{2}{\theta}\right)d\phi\right)^2+\frac{\Sigma}{\Delta_{r}}dr^2\\
    &+\frac{\sin^{2}{\theta}}{\Sigma}\left(a dt - \left(r^2+a^2+n^2-2anc\right)d\phi \right)^{2}+\Sigma d\theta^{2},    
\end{aligned}
\end{equation}
where

\begin{equation}
    \Delta_{r}= r^2-2mr+a^2-n^2,\:\Sigma \; = r^2 + (n+a\cos{\theta})^2.
\end{equation}
The Killing vectors for the spacetime are

\begin{equation}\label{killing1}
    \xi=\partial_{t},\:\chi=\partial_{\phi}.
\end{equation}

The Komar integrals use the symmetries inherent to the Killing vectors to find the associated conserved charges. Evaluating the integrals over the boundaries of the manifold, $M$, yields conserved quantities that we identify with the mass and appropriate thermodynamic charges and potentials. 

The following Killing vector generates a Killing horizon that coincides with the black hole horizon

\begin{equation}
    \xi+\Omega_{H}\chi
\end{equation}
where $\Omega_{H}$ is the angular velocity at the horizon. The generator of the horizon can be used to construct the following boundary integral
\begin{equation}
    0=\frac{1}{8\pi}\int_{\partial M} \star d\xi+\frac{\Omega_{H}}{8\pi}\int_{\partial M} \star d\chi.
\end{equation}

The correct identification of the spacelike boundaries is crucial. There are four boundaries: the black hole horizon($r=r_{h}$), the surface where ($r\rightarrow\infty$), and the two Misner strings. The Misner strings are approached by constructing tubes around the poles. The surfaces for the north and south strings, respectively, are defined by the hypersurfaces $\theta=\epsilon$ $(T_{+})$ and $\theta=\pi-\epsilon$ $(T_{-})$, taking the limit where $\epsilon \rightarrow 0$ \cite{bordo_misner_2019}. Taking the into account the orientation of the normal vector to each boundary, we find that we have 

\begin{equation}\label{Boundaries}
    \partial M =S_{r\rightarrow\infty}-S_{r_{h}}+ T_{+}-T_{-}.
\end{equation}
\noindent
Writing out the integrals over the boundaries explicitly, 

\begin{equation}
    \begin{aligned}
        0=& \frac{1}{8\pi}\int_{r\rightarrow\infty} \star d\xi-\frac{1}{8\pi}\int_{r_{h}} \star d\xi +\frac{1}{8\pi}(\int_{T_{+}} - \int_{T_{-}})\star d\xi\\
         &+\frac{\Omega_{H}}{8\pi}\int_{r\rightarrow\infty} \star d\chi-\frac{\Omega_{H}}{8\pi}\int_{r_{h}} \star d\chi+\frac{\Omega_{H}}{8\pi}(\int_{T_{+}} - \int_{T_{-}})\star d\chi.
    \end{aligned}
\end{equation}
The integral of $\star d\chi$ is broken down into two components where 
\begin{equation}
    \int_{T_{\pm}}\star d\chi=\Pi_{\pm}^{\infty}-\Pi_{\pm}^{r_{h}}.
\end{equation}
We rearrange the terms, grouping them into the corresponding thermodynamic terms

\begin{equation}
    \begin{aligned}\label{Kerr-NUT-Mink-Smarr-Grouped}
        \frac{-1}{8\pi}\int_{r\rightarrow\infty} \star d\xi =& 2\left(\frac{-1}{16\pi}\int_{r_{h}} \star d\xi +\Omega_{H}\star d\chi \right)+2\left(\frac{1}{16\pi}\int_{T_{+}} \star d\xi\right)\\
         &+2\left(\frac{-1}{16\pi}\int_{T_{-
         }} \star d\xi\right)+2\left(\frac{\Omega_{H}}{16\pi}\left(\int_{r\rightarrow\infty}\star d\chi +\Pi_{+}^{\infty}-\Pi_{-}^{\infty}\right) \right)\\
         &+2\left(\frac{\Omega_{H}}{16\pi}\left(\Pi_{-}^{r_{h}}-\Pi_{+}^{r_{h}}\right) \right).
    \end{aligned}
\end{equation}
\noindent
Replacing integrals with the associated thermodynamic quantities,
\begin{equation}\label{Minkowski-Charges}
    \begin{aligned}
        &M = \frac{-1}{8\pi}\int_{r\rightarrow\infty} \star d\xi,&\: 
        N_{+}\Phi_{N_{+}} =& \frac{1}{16\pi}\int_{T_{+}} \star d\xi,\\
        &TS = \frac{-1}{16\pi}\int_{r_{h}} \star d\xi +\Omega_{H}\star d\chi,&\:
        N_{-}\Phi_{N_{-}} =& \frac{-1}{16\pi} \int_{T_{-}}\star d\xi,\\
        &\Omega_{H} J_{bh} = \frac{\Omega_{H}}{16\pi}\left(\int_{r\rightarrow\infty}\star d\chi +\Pi_{+}^{\infty}-\Pi_{-}^{\infty}\right),&\:
        \Omega_{H} J_{s} =&\frac{\Omega_{H}}{16\pi}\left(\Pi_{-}^{r_{h}}-\Pi_{+}^{r_{h}}\right).
    \end{aligned}
\end{equation}
\noindent
Due to the integrals closing, the following the Smarr relation is manifestly satisfied,
\begin{equation}\label{Smarr-Minkowski}
        M = 2TS+2N_{+}\Phi_{N_{+}}+2N_{-}\Phi_{N_{-}}+2\Omega_{H}J_{bh}+2\Omega_{H}J_{s}.
\end{equation}
We assign two different nut charges and potentials. The nut charge $N_{+}$ corresponds to the contribution of the north string while $N_{-}$ represents the contribution of the south string. 

This choice is motivated by two reasons. First, it highlights that the nut charge vanishes as the each string vanishes. For example, when the north string vanishes we have $N_{+}=0$. Second, the given breakdown makes it more straightforward to compare our approach to the existing literature.

\noindent
The entropy is given by the Bekenstein-Hawking area law 
\begin{equation}
    S=\frac{A_{H}}{4}=\pi\left( r_{h}^{2}+a^{2}+n^{2}-2anc\right).
\end{equation}
Due to the dependence on $c$, there are two independent nut charges $N_{+}$ and $N_{-}$. The nut charges and their corresponding potentials are given by

\begin{equation}
    \begin{aligned}\label{Nut-charges-pm}
        N_{+}=n(1+c),\:\:
        \Phi_{N_{+}}=-\frac{n-a}{4r_{h}},\\
        N_{-}=n(1-c),\:\:
        \Phi_{N_{-}}=-\frac{n+a}{4r_{h}}.
    \end{aligned}
\end{equation}

Curiously, this identification of nut charges is proportional to the Noether charge entropies $\hat{N}^{'}_{\pm}$ when calculated between the horizon and some finite $R$. This was calculated in \cite{bordo_first_2019}. If we define a nut charge using $\hat{N}^{'}_{\pm}$ normalised per unit length, the nut charges \eqref{Nut-charges-pm} become related by a constant multiplicative factor such that $\hat{N}^{'}_{\pm}=2\pi N_{\pm}$. 

However, it is important to note that this argument is not the reason behind our charge identification. This choice is motivated by our aim to ease readability vis-\'a-vis other works in the literature. While we ultimately show that this choice is not necessarily lacking, we present a more appropriate and verifiably conserved choice of charges in the next section.

We differentiate between the black hole $J_{bh}$ and string $J_{s}$ contributions to the angular momentum. This distinction needs to be made to make sure the AdS thermodynamics hold. Separating them ensures that the approach is consistent with the generalised case while also serving the purpose of making comparisons to the existing literature easier. Interestingly, the angular momentum attributed to the black hole is equal to the angular momentum calculated using the surface charge method \cite{frodden_first_2022}. They are given by

\begin{equation}\label{angular-momnetum}
    J_{bh} = (a-3cn)m,\:\:\:\: J_{s}=\frac{n}{r_{h}}\left(an +c\left(r_{h}^{2}+acn-2n^{2}\right) \right).
\end{equation}
The angular velocity at the horizon is given by 
\begin{equation}
    \Omega_{H}=-\frac{g_{t\phi}}{g_{\phi\phi}}\Big\vert_{r=r_{h}}=\frac{a}{r_{h}^{2}+a^{2}+n^{2}-2acn}
\end{equation}
The thermodynamic mass is given by the mass parameter
\begin{equation}\label{thermodynamic-mass}
    M=m.
\end{equation}
These parameters satisfy the following first law
\begin{equation}\label{first-law-minkowksi_+_-}
    dU= T dS +\Phi_{N_{+}} dN_{+} +\Phi_{N_{-}} dN_{-} + \Omega_{H} dJ_{bh}+\Omega_{H} dJ_{s}.
\end{equation}
The internal energy is given by
\begin{equation}\label{internal-energy-pm}
    U=M-N_{+}\Phi_{+}-N_{-}\Phi_{-}.
\end{equation}

While the presented formulation differs from the works presented in the literature through the different choice of nut charges, it mirrors the literature in its attribution of a nut charge and potential to each string.

\section{Misner Strings and Nut Charge choices}

This section rethinks the attribution of nut charges, proving that the designation used in the previous section it not unique. Examining the choice of nut charges and its implications highlights the dissimilarity of our formulation to the other approaches in the literature. Before comparing the charges $N_{\pm}$ with alternative formulations, we consider how they are related to previous work that identifies the charges in a similar way.

We examine the nut charges as defined through \eqref{Nut-charges-pm} in the case where the rotation only due to $c$ or only due to $a$. If either is equal to zero, the designation of two different nut charges is not necessary. We propose that a formulation where the charges are $N_{n}=n$ and $N_{c}=cn$ and show that each charge is an independently conserved quantity. 

Taking the case where one of $a$ or $c$ vanishes $\left((a=0) \lor (c=0)\right)$, our results are consistent with earlier work \cite{awad_lorentzian_2022} where the total energy was given by
\begin{equation}
    U=M-N\Phi_{N},
\end{equation}
where
\begin{equation}\label{Previous-N-Phi_N}
    N=n,\:\:\Phi_{N}=-\frac{n}{2r_{h}}.
\end{equation}

This is interesting because it shows that there is a coupling between the traditional Kerr metric rotational contribution encoded in $a$ and the nut-related contribution represented by $c$. The separation into two distinct nut charges and potentials is only necessary if both $a\neq0$ and $c\neq0$. If either one vanishes, the contribution of the separated charges to the first law only depends on $n$. To see this, we write out the nut charge and associated potential for each case.

\noindent
The case of $c=0$ is straightforward as we have $N_{+}=N_{-}=n=N$. The contribution of the nut charges to the first law is given by
\begin{equation}
\begin{aligned}
    \left(\Phi_{N_{+}} dN_{+} +\Phi_{N_{-}} dN_{-}\right)\Big\vert_{c=0}
    &=\Phi_{N_{+}} dN +\Phi_{N_{-}} dN\\
    &=\left(\Phi_{N_{+}} +\Phi_{N_{-}} \right)dN 
    &=\left( \frac{-n}{2r_{h}}\right)dN. 
\end{aligned}
\end{equation}
This matches exactly with the values in \eqref{Previous-N-Phi_N}.

\noindent
For the case where $a=0$, 
\begin{equation}
    \begin{aligned}
        N_{+}=n(1+c),\:\:
        \Phi_{N_{+}}=-\frac{n}{4r_{h}}=\Phi^{'}_{N},\\
        N_{-}=n(1-c),\:\:
        \Phi_{N_{-}}=-\frac{n}{4r_{h}}=\Phi^{'}_{N}.
    \end{aligned}
\end{equation}
The contribution of the $c$-dependent portion on the first law is given by
\begin{equation}
\begin{aligned}\label{nut-charge-a=0}
    \left(\Phi_{N_{+}} dN_{+} +\Phi_{N_{-}} dN_{-}\right)\Big\vert_{a=0}
    &=\Phi^{'}_{N} d\left(n(1+c)\right) +\Phi^{'}_{N} d\left(n(1-c)\right)\\
    &=\Phi^{'}_{N}d\left(n(1+c)+n(1-c)\right)\\
    &=\Phi^{'}_{N}d\left(2n\right)\\
    &=2\Phi^{'}_{N}d\left(n\right)
\end{aligned}
\end{equation}
Taking $\Phi_{N}=2\Phi^{'}_{N}$ and $N=n$ matches exactly with the previous results \eqref{Previous-N-Phi_N}. 

More importantly, we highlight how the contributions can be completely separated into a nut charge term and a coupling term. This can be achieved by defining two charges $N_{n}$ and $N_{c}$ along with their respective potentials where $N_{n}$ does not depend on $c$. Using the following identification 
\begin{equation}\label{Nut-charge-seperated}
    N_{n}=\frac{N_{+}+N_{-}}{2}=n, \:\: N_{c}=\frac{N_{+}-N_{-}}{2}=cn,
\end{equation}\label{Nut-potential-seperated}
where the corresponding potentials are given by 
\begin{equation}
    \Phi_{N_{n}}=\frac{-n}{2r_{h}}, \:\: \Phi_{N_{c}}=\frac{a}{2r_{h}}.
\end{equation}
Using these identifications we can write the following first law 

\begin{equation}\label{first-law-minkowksi_c_n}
    dU= T dS +\Phi_{N_{n}} dN_{n} +\Phi_{N_{c}} dN_{c} + \Omega_{H} dJ_{bh}+\Omega_{H} dJ_{s}, 
\end{equation}
where 
\begin{equation}\label{U-minkowksi_n_c}
    U= M-N_{n}\Phi_{N_{n}} -N_{c}\Phi_{N_{c}} .
\end{equation}

The ability to separate them in this manner is quite important. To consistently define a thermodynamic charge, we must ensure that it is a conserved quantity. Accordingly, it is important to demonstrate that the decomposition \eqref{Nut-charge-seperated} consists of fixed thermodynamic quantities. The charge $N_{n}=n$ is fixed through the gravito-magnetic duality that exists with $m$. Accordingly, we must show that $cn$ is a conserved quantity. 

This is straightforward for the case of $a=0$ since the conserved black hole angular momentum would equal $J_{bh}=-3cnm$. In this case, we can show that $c$ is independently conserved. The parameter $m$ is fixed by fixing the thermodynamic mass $M$. Since $J_{bh}$ is a conserved thermodynamic quantity that is independent of $M$, it follows that the charge $cn$ is conserved.

When $a\neq0$ it is not as straightforward to verify that $cn$ is a conserved quantity. For this case, the conserved angular momentum is $J_{bh}=\left(a-3cn\right)m$. Observing that $m$ is a fixed parameter only implies that $a-3cn$ is conserved. Accordingly, we must find another way to show that $cn$ is a conserved quantity.

To demonstrate that $cn$ corresponds to a conserved thermodynamic quantity, we consider the Hodge dual to the two-form associated with the angular momentum; $\star\left(\star d\chi\right)$. A straight forward calculation shows that, similar to the dual of the mass, the associated integral closes over the boundaries. Evaluating this integral over the surface where $r\rightarrow\infty$ yields
\begin{equation}\label{cn^2-midtext}
    \frac{-1}{8\pi}\int_{S^{2}_{r\rightarrow\infty}}  d\chi=2cn^{2}=\left(cn\right)\cdot\left(2n\right).
\end{equation}
Since the conserved quantity $2cn^{2}$ is the product of $cn$ with $2n$ and is fixed independently from $n$, it follows that $cn$ is conserved. This observation is consistent with the fact that the thermodynamic charges $N_{+}$ and $N_{-}$ satisfy the first law \eqref{first-law-minkowksi_+_-}. Since the linear combination of any two valid thermodynamic charges results in a consistent first law, this is not surprising. 

It is important to note that this result \eqref{cn^2-midtext} holds regardless of the existence of strings as physical boundaries. While the integral for $\star d\chi$ requires the addition of the string contributions at infinity to be regularised, the integral for $d\chi$ does not.

Thus far, our discussion has served two purposes. First, it has shown that the identification of charges in \eqref{Nut-charges-pm} is not a problematic assignment. Second, it has opened the door to a different way of interpreting the nut charge contribution to the thermodynamics. 

While we can consider the nut charges to be contributions of two different strings, we can also use an $n$-related term and a $cn$ term. The first views the Misner strings as independent contributors to the thermodynamics which highlights how the contribution of the integral over a hidden string is always zero (ex. $N_{+}=0$ for $c=-1$). The second focuses on the independent contributions of $n$ and $cn$. It highlights how the work term is due to some coupling with the rotational parameter $a$ and how the original nut contribution is independent of the value of $c$.

The only reason for the choice $cn$ is convention. The value of $c=-1$ ensures that north string is hidden while $c=1$ hides the south string. It is important to note that the addition of $c$ is done through the large coordinate transformation $t\rightarrow t+2nc\phi$. Another influence is the work presented by Cl\'ement et al. where they show that the Taub-NUT spacetime does not contain any closed timelike or null geodesics for $\vert c\vert\leq 1$. 

In principle, this parameter does not need to depend on $n$. The identification of the $cn$ term with a parameter that does not depend on $n$ is already present in the literature. For example, the authors in \cite{bordo_first_2019} used the identification $cn=s$ to construct their thermodynamics. However, they did not treat it as an independent thermodynamic charge. 

\noindent
Taking the choice of $N_{c}=cn=s$ we can verify that our first law \eqref{first-law-minkowksi_c_n} still holds 
\begin{equation}
    dU= T dS +\Phi_{N_{n}} dN_{n} +\Phi_{s} ds + \Omega_{H} dJ_{bh}+\Omega_{H} dJ_{s}.
\end{equation}

Lastly, we wish to highlight that our quantities are all expressed in terms of parameters that are demonstrably fixed due through conservation of physical quantities. As shown above, both $n$ and $cn$ are independently conserved. Consequently, we need to show that the two other parameters $r_{h}$ and $a$ are fixed by charge conservation.

For the parameter $a$, we note that the mass \eqref{thermodynamic-mass} and the angular momentum \eqref{angular-momnetum} are both independently conserved. Due to both quantities being independently conserved and  the angular momentum $J_{bh}$ being equal to $(a-3cn)m$, it follows that $(a-3cn)$ is a conserved quantity. By noting the independent conservation of $cn$ we deduce that the parameter $a$ is fixed. 

Finally, the conservation of the mass fixes the horizon radius $r_{h}$. The mass term can be written in terms in terms of the given parameters as $m(r_{h},a,n)$. Given that mass conservation does not depend on $a$ or $n$, it is clear that $r_{h}$ is a fixed parameter. 
\section{Kerr-NUT Thermodynamics from the Action}
In this section, we use the Euclidean action $I$ to calculate the Gibbs energy $G$. The differential form of the Gibbs energy can be used to obtain the first law through the application of Legendre transforms. We show that the assignment of charges in \eqref{Nut-charge-seperated} satisfies the differential form of the Gibbs energy equation.

The action is made up of two terms: The first is the Einstein-Hilbert term which is an integral over the bulk $M$, while the second is the Gibbons-Hawking Boundary term that depends on the extrinsic curvature $K$ and the determinant of induced metric $h_{ab}$ at the boundary $\partial M$

\begin{equation}
    \begin{aligned}\label{action-Minkowski}
        &I=I_{EH}+I_{GH},\\
        &I_{EH}=\int_{M}\sqrt{-g}R,\\
        &I_{GH}=\int_{\partial M}\sqrt{-h}K.
    \end{aligned}
\end{equation}
The action is equal to 

\begin{equation}
    I=\beta\frac{m}{2}.
\end{equation}
The Gibbs energy $G$ is given by $\frac{I}{\beta}$. Writing $G$ explicitly, 
\begin{equation}
    G=\frac{r_{h}^2-n^2+a^2}{4r_{h}}.
\end{equation}

We define the thermodynamic parameters as we did in the previous section. Taking the partial derivatives with respect to the fixed quantities we find:
\begin{equation}
    \begin{aligned}
        \left(\frac{\partial G}{\partial T}\right)\Bigg|_{(N_{n},N_{c},\Omega_{H})}\:= -S,&
         \left(\frac{\partial G}{\partial N_{n}}\right)\Bigg|_{(T,N_{c},\Omega_{H})}\:=\Phi_{N_{n}}, \\
         \left(\frac{\partial G}{\partial N_{c}}\right)\Bigg|_{(T,N_{n},\Omega_{H})}\:=\Phi_{N_{c}}, &
         \left(\frac{\partial G}{\partial \Omega_{H}}\right)\Bigg|_{(T,N_{n},N_{c})}\:=-J.
    \end{aligned}
\end{equation}

\noindent
Accordingly, the differential relation takes the form

\begin{equation}\label{Gibbs-Minkowski-pm}
     dG = -S dT +\Phi_{N_{n}} dN_{n} +\Phi_{N_{c}} dN_{c} - J d\Omega_H.
\end{equation}
The thermodynamic quantities \eqref{Minkowski-Charges} satisfy the Gibbs-Duhem relation: 

\begin{equation}\label{Gibbs-scaling}
    \frac{I}{\beta}=M-N_{n}\Phi_{N_{n}}-N_{c}\Phi_{N_{c}}-TS-J\Omega_{H}
\end{equation}
This can be used to obtain the internal energy of the spacetime which is related to the Gibbs energy through the 
\begin{equation}
    U=G+TS+J\Omega_{H},
\end{equation}
where the fixing of $\beta$ and $\Omega_{h}$ in the Euclidean path integral determines the Legendre transform.
It is straightforward to verify that the charges $N_{+}$ and $N_{-}$ also satisfy the differential Gibbs relation. 

\begin{equation}\label{Gibbs-Minkowski-nc}
     dG = -S dT +\Phi_{N_{+}} dN_{+} +\Phi_{N_{-}} dN_{-} - J d\Omega_H.
\end{equation}
The scaling relations \eqref{Smarr-Minkowski} and \eqref{Gibbs-scaling} are also satisfied by either choice since 
\begin{equation}
    N_{+}\Phi_{N_{+}} +N_{-}\Phi_{N_{-}}=N_{n}\Phi_{N_{n}} +N_{c}\Phi_{N_{c}}.
\end{equation}
\section{Conclusion}
In this work, we constructed the thermodynamics for the Kerr-NUT solution, where we used conserved charges that differ from the identifications present in the literature. We showed that, in addition to the traditional conserved dual mass, a conserved dual angular momentum is also present. This additional conserved quantity allows a well-motivated identification of our charges such that all the independent thermodynamic parameters are fixed as a direct consequence of charge conservation. 

The first law is presented in terms of two different sets of nut charges. One set is given by $N_{n}$ and $N_{c}$ and is written directly in terms of the conserved quantities; this is novel. The other is given by the $N_{+}$ and $N_{-}$ where the geometric contribution of each string is independently highlighted. Both sets are related through a linear transformation, and thus produce functionally equivalent thermodynamics. 

Furthermore, we verified that the thermodynamic contribution of the additional charge is due to a coupling between the Taub-NUT related rotation encoded in $cn$ and the Kerr rotation parameter $a$. The additional charge $N_{c}$ vanishes when $cn=0$ while its corresponding potential $\Phi_{N_{c}}$ dies for $a=0$. Accordingly, this contribution only arises when the two sources of overall angular momentum exist.

We used two different methods to construct our thermodynamic formulation. First, we demonstrated that a geometric calculation of the charges using Komar integrals results in a Smarr relation whose charges satisfy the first law of thermodynamics. Second, the Euclidean action was used to obtain the Gibbs energy which was used to show that the differential Gibbs energy equation is satisfied by the given thermodynamic parameters. We also show that the charges obey the Gibbs-Duhem relation. 

This work can be extended in several ways in the future. For example, the thermodynamics of the Kerr-NUT-adS solution with $cn\neq 0$ are yet to be satisfied in the literature. Extending our analysis to this case might lead to the appropriate identification of charges. While the phase structure has been studied for the $cn=0$ case \cite{bordo_thermodynamics_2020}, the case where it does not vanish has not been studied. Given that the presence of $n$ affects the distribution of the electromagnetic charges, it might be possible for $cn$ to have a similar effect. For 
example, if one of the electromagnetic charges or potentials explicitly depends on $cn$, there will be an ensemble where its value could have a direct effect on the phase structure.

\section{Acknowledgements}
The work of AA is partially supported by the Science, Technology \& Innovation Funding Authority (STDF) under grant number 50806.

The work of AA and MT is partially supported by the Science, Technology \& Innovation Funding Authority (STDF) under grant number 53090.
\printbibliography
\end{document}